%
%
%
\def\unredoffs{} \def\redoffs{\voffset=-.31truein\hoffset=-.48truein}
\def\speclscape{}

%
%
%
%
%
\newbox\leftpage \newdimen\fullhsize \newdimen\hstitle \newdimen\hsbody
\tolerance=1000\hfuzz=2pt
\catcode`\@=11 
\ifx\hyperdef\UNd@FiNeD\def\hyperdef#1#2#3#4{#4}\def\hyperref#1#2#3#4{#4}\fi
\def\bigans{b }
\def\answ{b }
%
\ifx\answ\bigans\message{(This will come out unreduced.}
\magnification=1200\unredoffs\baselineskip=16pt plus 2pt minus 1pt
\hsbody=\hsize \hstitle=\hsize 
\else\message{(This will be reduced.} \let\l@r=L
\magnification=1000\baselineskip=16pt plus 2pt minus 1pt \vsize=7truein
\redoffs \hstitle=8truein\hsbody=4.75truein\fullhsize=10truein\hsize=\hsbody
\output={\ifnum\pageno=0 
  \shipout\vbox{\speclscape{\hsize\fullhsize\makeheadline}
    \hbox to \fullhsize{\hfill\pagebody\hfill}}\advancepageno
  \else
  \almostshipout{\leftline{\vbox{\pagebody\makefootline}}}\advancepageno
  \fi}
\def\almostshipout#1{\if L\l@r \count1=1 \message{[\the\count0.\the\count1]}
      \global\setbox\leftpage=#1 \global\let\l@r=R
 \else \count1=2
  \shipout\vbox{\speclscape{\hsize\fullhsize\makeheadline}
      \hbox to\fullhsize{\box\leftpage\hfil#1}}  \global\let\l@r=L\fi}
\fi
%
\newcount\yearltd\yearltd=\year\advance\yearltd by -2000

\def\Title#1#2{\nopagenumbers\abstractfont\hsize=\hstitle\rightline{#1}%
\vskip 1in\centerline{\titlefont #2}\abstractfont\vskip .5in\pageno=0}
\def\Date#1{\vfill\leftline{#1}\tenpoint\supereject\global\hsize=\hsbody%
\footline={\hss\tenrm\hyperdef\hypernoname{page}\folio\folio\hss}}%
%

\def\draftmode{\message{ DRAFTMODE }\def\draftdate{{\rm preliminary draft:
\number\month/\number\day/\number\yearltd\ \ \hourmin}}%
\headline={\hfil\draftdate}\writelabels\baselineskip=20pt plus 2pt minus 2pt
 {\count255=\time\divide\count255 by 60 \xdef\hourmin{\number\count255}
  \multiply\count255 by-60\advance\count255 by\time
  \xdef\hourmin{\hourmin:\ifnum\count255<10 0\fi\the\count255}}}
\def\nolabels{\def\wrlabeL##1{}\def\eqlabeL##1{}\def\reflabeL##1{}}
\def\writelabels{\def\wrlabeL##1{\leavevmode\vadjust{\rlap{\smash%
{\line{{\escapechar=` \hfill\rlap{\sevenrm\hskip.03in\string##1}}}}}}}%
\def\eqlabeL##1{{\escapechar-1\rlap{\sevenrm\hskip.05in\string##1}}}%
\def\reflabeL##1{\noexpand\llap{\noexpand\sevenrm\string\string\string##1}}}
\nolabels
%
\global\newcount\secno \global\secno=0
\global\newcount\meqno \global\meqno=1
\def\s@csym{}
\def\newsec#1{\global\advance\secno by1%
{\toks0{#1}\message{(\the\secno. \the\toks0)}}%
\global\subsecno=0\eqnres@t\let\s@csym\secsym\xdef\secn@m{\the\secno}\noindent
{\bf\hyperdef\hypernoname{section}{\the\secno}{\the\secno.} #1}%
\writetoca{{\string\hyperref{}{section}{\the\secno}{\the\secno.}} {#1}}%
\par\nobreak\medskip\nobreak}
\def\eqnres@t{\xdef\secsym{\the\secno.}\global\meqno=1\bigbreak\bigskip}
\def\sequentialequations{\def\eqnres@t{\bigbreak}}\xdef\secsym{}
\global\newcount\subsecno \global\subsecno=0
\def\subsec#1{\global\advance\subsecno by1%
{\toks0{#1}\message{(\s@csym\the\subsecno. \the\toks0)}}%
\ifnum\lastpenalty>9000\else\bigbreak\fi
\noindent{\it\hyperdef\hypernoname{subsection}{\secn@m.\the\subsecno}%
{\secn@m.\the\subsecno.} #1}\writetoca{\string\quad
{\string\hyperref{}{subsection}{\secn@m.\the\subsecno}{\secn@m.\the\subsecno.}}
{#1}}\par\nobreak\medskip\nobreak}
\def\appendix#1#2{\global\meqno=1\global\subsecno=0\xdef\secsym{\hbox{#1.}}%
\bigbreak\bigskip\noindent{\bf Appendix \hyperdef\hypernoname{appendix}{#1}%
{#1.} #2}{\toks0{(#1. #2)}\message{\the\toks0}}%
\xdef\s@csym{#1.}\xdef\secn@m{#1}%
\writetoca{\string\hyperref{}{appendix}{#1}{Appendix {#1.}} {#2}}%
\par\nobreak\medskip\nobreak}
%
%
\def\checkm@de#1#2{\ifmmode{\def\f@rst##1{##1}\hyperdef\hypernoname{equation}%
{#1}{#2}}\else\hyperref{}{equation}{#1}{#2}\fi}
\def\eqnn#1{\DefWarn#1\xdef #1{(\noexpand\relax\noexpand\checkm@de%
{\s@csym\the\meqno}{\secsym\the\meqno})}%
\wrlabeL#1\writedef{#1\leftbracket#1}\global\advance\meqno by1}
\def\f@rst#1{\c@t#1a\em@ark}\def\c@t#1#2\em@ark{#1}
\def\eqna#1{\DefWarn#1\wrlabeL{#1$\{\}$}%
\xdef #1##1{(\noexpand\relax\noexpand\checkm@de%
{\s@csym\the\meqno\noexpand\f@rst{##1}}{\hbox{$\secsym\the\meqno##1$}})}
\writedef{#1\numbersign1\leftbracket#1{\numbersign1}}\global\advance\meqno by1}
\def\eqn#1#2{\DefWarn#1%
\xdef #1{(\noexpand\hyperref{}{equation}{\s@csym\the\meqno}%
{\secsym\the\meqno})}$$#2\eqno(\hyperdef\hypernoname{equation}%
{\s@csym\the\meqno}{\secsym\the\meqno})\eqlabeL#1$$%
\writedef{#1\leftbracket#1}\global\advance\meqno by1}
\def\xeqn{\expandafter\xe@n}\def\xe@n(#1){#1}
\def\xeqna#1{\expandafter\xe@n#1}
\def\eqns#1{(\e@ns #1{\hbox{}})}
\def\e@ns#1{\ifx\UNd@FiNeD#1\message{eqnlabel \string#1 is undefined.}%
\xdef#1{(?.?)}\fi{\let\hyperref=\relax\xdef\next{#1}}%
\ifx\next\em@rk\def\next{}\else%
\ifx\next#1\xeqn#1\else\def\n@xt{#1}\ifx\n@xt\next#1\else\xeqna#1\fi
\fi\let\next=\e@ns\fi\next}

\def\DefWarn#1{\ifx\UNd@FiNeD#1\else
\immediate\write16{*** WARNING: the label \string#1 is already defined ***}\fi}
%
\newskip\footskip\footskip14pt plus 1pt minus 1pt 
\def\footnotefont{\ninepoint}\def\f@t#1{\footnotefont #1\@foot}
\def\f@@t{\baselineskip\footskip\bgroup\footnotefont\aftergroup\@foot\let\next}
\setbox\strutbox=\hbox{\vrule height9.5pt depth4.5pt width0pt}
\global\newcount\ftno \global\ftno=0
\def\foot{\global\advance\ftno by1\def\foot@rg{\hyperref{}{footnote}%
{\the\ftno}{\the\ftno}\xdef\foot@rg{\noexpand\hyperdef\noexpand\hypernoname%
{footnote}{\the\ftno}{\the\ftno}}}\footnote{$^{\foot@rg}$}}
%
\newwrite\ftfile
\def\footend{\def\foot{\global\advance\ftno by1\chardef\wfile=\ftfile
\hyperref{}{footnote}{\the\ftno}{$^{\the\ftno}$}%
\ifnum\ftno=1\immediate\openout\ftfile=\jobname.fts\fi%
\immediate\write\ftfile{\noexpand\smallskip%
\noexpand\item{\noexpand\hyperdef\noexpand\hypernoname{footnote}
{\the\ftno}{f\the\ftno}:\ }\pctsign}\findarg}%
\def\footatend{\vfill\eject\immediate\closeout\ftfile{\parindent=20pt
\centerline{\bf Footnotes}\nobreak\bigskip\input \jobname.fts }}}
\def\footatend{}
%
%
\global\newcount\refno \global\refno=1
\newwrite\rfile
\def\ref{[\hyperref{}{reference}{\the\refno}{\the\refno}]\nref}
\def\nref#1{\DefWarn#1%
\xdef#1{[\noexpand\hyperref{}{reference}{\the\refno}{\the\refno}]}%
\writedef{#1\leftbracket#1}%
\ifnum\refno=1\immediate\openout\rfile=\jobname.refs\fi
\chardef\wfile=\rfile\immediate\write\rfile{\noexpand\item{[\noexpand\hyperdef%
\noexpand\hypernoname{reference}{\the\refno}{\the\refno}]\ }%
\reflabeL{#1\hskip.31in}\pctsign}\global\advance\refno by1\findarg}
\def\findarg#1#{\begingroup\obeylines\newlinechar=`\^^M\pass@rg}
{\obeylines\gdef\pass@rg#1{\writ@line\relax #1^^M\hbox{}^^M}%
\gdef\writ@line#1^^M{\expandafter\toks0\expandafter{\striprel@x #1}%
\edef\next{\the\toks0}\ifx\next\em@rk\let\next=\endgroup\else\ifx\next\empty%
\else\immediate\write\wfile{\the\toks0}\fi\let\next=\writ@line\fi\next\relax}}
\def\striprel@x#1{} \def\em@rk{\hbox{}}
\def\lref{\begingroup\obeylines\lr@f}
\def\lr@f#1#2{\DefWarn#1\gdef#1{\let#1=\UNd@FiNeD\ref#1{#2}}\endgroup\unskip}

\def\addref#1{\immediate\write\rfile{\noexpand\item{}#1}} 
\def\listrefs{\footatend\vfill\supereject\immediate\closeout\rfile\writestoppt
\baselineskip=\footskip\centerline{{\bf References}}\bigskip{\parindent=20pt%
\frenchspacing\escapechar=` \input \jobname.refs\vfill\eject}\nonfrenchspacing}
\def\startrefs#1{\immediate\openout\rfile=\jobname.refs\refno=#1}
\def\xref{\expandafter\xr@f}\def\xr@f[#1]{#1}
\def\refs#1{\count255=1[\r@fs #1{\hbox{}}]}
\def\r@fs#1{\ifx\UNd@FiNeD#1\message{reflabel \string#1 is undefined.}%
\nref#1{need to supply reference \string#1.}\fi%
\vphantom{\hphantom{#1}}{\let\hyperref=\relax\xdef\next{#1}}%
\ifx\next\em@rk\def\next{}%
\else\ifx\next#1\ifodd\count255\relax\xref#1\count255=0\fi%
\else#1\count255=1\fi\let\next=\r@fs\fi\next}
%

%
\newwrite\ffile\global\newcount\figno \global\figno=1
\def\fig{fig.~\hyperref{}{figure}{\the\figno}{\the\figno}\nfig}
\def\nfig#1{\DefWarn#1%
\xdef#1{fig.~\noexpand\hyperref{}{figure}{\the\figno}{\the\figno}}%
\writedef{#1\leftbracket fig.\noexpand~\xfig#1}%
\ifnum\figno=1\immediate\openout\ffile=\jobname.figs\fi\chardef\wfile=\ffile%
{\let\hyperref=\relax
\immediate\write\ffile{\noexpand\medskip\noexpand\item{Fig.\ %
\noexpand\hyperdef\noexpand\hypernoname{figure}{\the\figno}{\the\figno}. }
\reflabeL{#1\hskip.55in}\pctsign}}\global\advance\figno by1\findarg}
\def\listfigs{\vfill\eject\immediate\closeout\ffile{\parindent40pt
\baselineskip14pt\centerline{{\bf Figure Captions}}\nobreak\medskip
\escapechar=` \input \jobname.figs\vfill\eject}}
\def\xfig{\expandafter\xf@g}\def\xf@g fig.\penalty\@M\ {}
\def\figs#1{figs.~\f@gs #1{\hbox{}}}
\def\f@gs#1{{\let\hyperref=\relax\xdef\next{#1}}\ifx\next\em@rk\def\next{}\else
\ifx\next#1\xfig #1\else#1\fi\let\next=\f@gs\fi\next}
\def\figin{\epsfcheck\figin}\def\figins{\epsfcheck\figins}
\def\epsfcheck{\ifx\epsfbox\UNd@FiNeD
\message{(NO epsf.tex, FIGURES WILL BE IGNORED)}
\gdef\figin##1{\vskip2in}\gdef\figins##1{\hskip.5in}
\else\message{(FIGURES WILL BE INCLUDED)}%
\gdef\figin##1{##1}\gdef\figins##1{##1}\fi}
\def\DefWarn#1{}
\def\figinsert{\goodbreak\midinsert}
\def\ifig#1#2#3{\DefWarn#1\xdef#1{fig.~\noexpand\hyperref{}{figure}%
{\the\figno}{\the\figno}}\writedef{#1\leftbracket fig.\noexpand~\xfig#1}%
\figinsert\figin{\centerline{#3}}\medskip\centerline{\vbox{\baselineskip12pt
\advance\hsize by -1truein\noindent\wrlabeL{#1=#1}\footnotefont%
{\bf Fig.~\hyperdef\hypernoname{figure}{\the\figno}{\the\figno}:} #2}}
\bigskip\endinsert\global\advance\figno by1}
\newwrite\lfile
{\escapechar-1\xdef\pctsign{\string\%}\xdef\leftbracket{\string\{}
\xdef\rightbracket{\string\}}\xdef\numbersign{\string\#}}
\def\writedefs{\immediate\openout\lfile=\jobname.defs \def\writedef##1{%
{\let\hyperref=\relax\let\hyperdef=\relax\let\hypernoname=\relax
 \immediate\write\lfile{\string\def\string##1\rightbracket}}}}%
\def\writestop{\def\writestoppt{\immediate\write\lfile{\string\pageno
 \the\pageno\string\startrefs\leftbracket\the\refno\rightbracket
 \string\def\string\secsym\leftbracket\secsym\rightbracket
 \string\secno\the\secno\string\meqno\the\meqno}\immediate\closeout\lfile}}
\def\writestoppt{}\def\writedef#1{}
\def\seclab#1{\DefWarn#1%
\xdef #1{\noexpand\hyperref{}{section}{\the\secno}{\the\secno}}%
\writedef{#1\leftbracket#1}\wrlabeL{#1=#1}}
\def\subseclab#1{\DefWarn#1%
\xdef #1{\noexpand\hyperref{}{subsection}{\secn@m.\the\subsecno}%
{\secn@m.\the\subsecno}}\writedef{#1\leftbracket#1}\wrlabeL{#1=#1}}
\def\applab#1{\DefWarn#1%
\xdef #1{\noexpand\hyperref{}{appendix}{\secn@m}{\secn@m}}%
\writedef{#1\leftbracket#1}\wrlabeL{#1=#1}}
\newwrite\tfile \def\writetoca#1{}
\def\leaderfill{\leaders\hbox to 1em{\hss.\hss}\hfill}
\def\writetoc{\immediate\openout\tfile=\jobname.toc
   \def\writetoca##1{{\edef\next{\write\tfile{\noindent ##1
   \string\leaderfill {\string\hyperref{}{page}{\noexpand\number\pageno}%
                       {\noexpand\number\pageno}} \par}}\next}}}
\newread\ch@ckfile
\def\listtoc{\immediate\closeout\tfile\immediate\openin\ch@ckfile=\jobname.toc
\ifeof\ch@ckfile\message{no file \jobname.toc, no table of contents this pass}%
\else\closein\ch@ckfile\centerline{\bf Contents}\nobreak\medskip%
{\baselineskip=12pt\footnotefont\parskip=0pt\catcode`\@=11\input\jobname.toc
\catcode`\@=12\bigbreak\bigskip}\fi}
\catcode`\@=12 
%
\edef\tfontsize{\ifx\answ\bigans scaled\magstep3\else scaled\magstep4\fi}
\font\titlerm=cmr10 \tfontsize \font\titlerms=cmr7 \tfontsize
\font\titlermss=cmr5 \tfontsize \font\titlei=cmmi10 \tfontsize
\font\titleis=cmmi7 \tfontsize \font\titleiss=cmmi5 \tfontsize
\font\titlesy=cmsy10 \tfontsize \font\titlesys=cmsy7 \tfontsize
\font\titlesyss=cmsy5 \tfontsize \font\titleit=cmti10 \tfontsize
\skewchar\titlei='177 \skewchar\titleis='177 \skewchar\titleiss='177
\skewchar\titlesy='60 \skewchar\titlesys='60 \skewchar\titlesyss='60
\def\titlefont{\def\rm{\fam0\titlerm}
\textfont0=\titlerm \scriptfont0=\titlerms \scriptscriptfont0=\titlermss
\textfont1=\titlei \scriptfont1=\titleis \scriptscriptfont1=\titleiss
\textfont2=\titlesy \scriptfont2=\titlesys \scriptscriptfont2=\titlesyss
\textfont\itfam=\titleit \def\it{\fam\itfam\titleit}\rm}
 \ifx\answ\bigans\else scaled\magstep1\fi
\ifx\answ\bigans\def\abstractfont{\tenpoint}\else
\font\absit=cmti10 scaled \magstep1
\font\abssl=cmsl10 scaled \magstep1
\font\absrm=cmr10 scaled\magstep1 \font\absrms=cmr7 scaled\magstep1
\font\absrmss=cmr5 scaled\magstep1 \font\absi=cmmi10 scaled\magstep1
\font\absis=cmmi7 scaled\magstep1 \font\absiss=cmmi5 scaled\magstep1
\font\abssy=cmsy10 scaled\magstep1 \font\abssys=cmsy7 scaled\magstep1
\font\abssyss=cmsy5 scaled\magstep1 \font\absbf=cmbx10 scaled\magstep1
\skewchar\absi='177 \skewchar\absis='177 \skewchar\absiss='177
\skewchar\abssy='60 \skewchar\abssys='60 \skewchar\abssyss='60
\def\abstractfont{\def\rm{\fam0\absrm}
\textfont0=\absrm \scriptfont0=\absrms \scriptscriptfont0=\absrmss
\textfont1=\absi \scriptfont1=\absis \scriptscriptfont1=\absiss
\textfont2=\abssy \scriptfont2=\abssys \scriptscriptfont2=\abssyss
\textfont\itfam=\absit \def\it{\fam\itfam\absit}\def\footnotefont{\tenpoint}%
\textfont\slfam=\abssl \def\sl{\fam\slfam\abssl}%
\textfont\bffam=\absbf \def\bf
{\fam\bffam\absbf}\rm}\fi
\def\tenpoint{\def\rm{\fam0\tenrm}
\textfont0=\tenrm \scriptfont0=\sevenrm \scriptscriptfont0=\fiverm
\textfont1=\teni  \scriptfont1=\seveni  \scriptscriptfont1=\fivei
\textfont2=\tensy \scriptfont2=\sevensy \scriptscriptfont2=\fivesy
\textfont\itfam=\tenit \def\it{\fam\itfam\tenit}\def\footnotefont{\ninepoint}%
\textfont\bffam=\tenbf \def\bf{\fam\bffam\tenbf}\def\sl{\fam\slfam\tensl}\rm}
\font\ninerm=cmr9 \font\sixrm=cmr6 \font\ninei=cmmi9 \font\sixi=cmmi6
\font\ninesy=cmsy9 \font\sixsy=cmsy6 \font\ninebf=cmbx9
\font\nineit=cmti9 \font\ninesl=cmsl9 \skewchar\ninei='177
\skewchar\sixi='177 \skewchar\ninesy='60 \skewchar\sixsy='60
\def\ninepoint{\def\rm{\fam0\ninerm}
\textfont0=\ninerm \scriptfont0=\sixrm \scriptscriptfont0=\fiverm
\textfont1=\ninei \scriptfont1=\sixi \scriptscriptfont1=\fivei
\textfont2=\ninesy \scriptfont2=\sixsy \scriptscriptfont2=\fivesy
\textfont\itfam=\ninei \def\it{\fam\itfam\nineit}\def\sl{\fam\slfam\ninesl}%
\textfont\bffam=\ninebf \def\bf{\fam\bffam\ninebf}\rm}
%
%
\def\noblackbox{\overfullrule=0pt}
\hyphenation{anom-aly anom-alies coun-ter-term coun-ter-terms}
\def\inv{^{\raise.15ex\hbox{${\scriptscriptstyle -}$}\kern-.05em 1}}

\def\Dsl{\,\raise.15ex\hbox{/}\mkern-13.5mu D} 
\def\dsl{\raise.15ex\hbox{/}\kern-.57em\partial}

\def\lspace{\ifx\answ\bigans{}\else\qquad\fi}
\def\lbspace{\ifx\answ\bigans{}\else\hskip-.2in\fi} 

\def\boxeqn#1{\vcenter{\vbox{\hrule\hbox{\vrule\kern3pt\vbox{\kern3pt
	\hbox{${\displaystyle #1}$}\kern3pt}\kern3pt\vrule}\hrule}}}
\def\mbox#1#2{\vcenter{\hrule \hbox{\vrule height#2in
		\kern#1in \vrule} \hrule}}  

\def\grad#1{\,\nabla\!_{{#1}}\,}

\def\darr#1{\raise1.5ex\hbox{$\leftrightarrow$}\mkern-16.5mu #1}

\def\roughly#1{\raise.3ex\hbox{$#1$\kern-.75em\lower1ex\hbox{$\sim$}}}

\input amssym
\input epsf

\def\IZ{\relax\ifmmode\mathchoice
{\hbox{\cmss Z\kern-.4em Z}}{\hbox{\cmss Z\kern-.4em Z}} {\lower.9pt\hbox{\cmsss Z\kern-.4em Z}}
{\lower1.2pt\hbox{\cmsss Z\kern-.4em Z}}\else{\cmss Z\kern-.4em Z}\fi}

\newif\ifdraft\draftfalse
\newif\ifinter\interfalse
\ifdraft\draftmode\else\interfalse\fi
\def\journal#1&#2(#3){\unskip, \sl #1\ \bf #2 \rm(19#3) }
\def\andjournal#1&#2(#3){\sl #1~\bf #2 \rm (19#3) }

\def\frac#1#2{{#1\over#2}}

\def\ds{\displaystyle}

\def\inbar{\,\vrule height1.5ex width.4pt depth0pt}
\def\IC{\relax\hbox{$\inbar\kern-.3em{\rm C}$}}
\def\IR{\relax{\rm I\kern-.18em R}}
\def\IP{\relax{\rm I\kern-.18em P}}

%
%


%
\catcode`\@=11
\def\slash#1{\mathord{\mathpalette\c@ncel{#1}}}
\overfullrule=0pt

\def\CC{{\cal C}}

\def\Z{\hbox{$\bb Z$}}
\def\R{\hbox{$\bb R$}}

\def\underrel#1\over#2{\mathrel{\mathop{\kern\z@#1}\limits_{#2}}}

\catcode`\@=12


%


\def\[{[}
\def\]{]}

\def\comment#1{ }

%
\def\draftnote#1{\ifdraft{\baselineskip2ex
                 \vbox{\kern1em\hrule\hbox{\vrule\kern1em\vbox{\kern1ex
                 \noindent \underbar{NOTE}: #1
             \vskip1ex}\kern1em\vrule}\hrule}}\fi}
\def\internote#1{\ifinter{\baselineskip2ex
                 \vbox{\kern1em\hrule\hbox{\vrule\kern1em\vbox{\kern1ex
                 \noindent \underbar{Internal Note}: #1
             \vskip1ex}\kern1em\vrule}\hrule}}\fi}

%

%
%

%

\def\inv{^{-1}}



\def\b{\beta}


\def\bb{
\font\tenmsb=msbm10
\font\sevenmsb=msbm7
\font\fivemsb=msbm5
\textfont1=\tenmsb
\scriptfont1=\sevenmsb
\scriptscriptfont1=\fivemsb
}





\def\bar{\overline}
\def\b{\bar}
\def\bsq#1{{{\b{#1}}^{\lower 2.5pt\hbox{$\scriptstyle 2$}}}}
\def\bexp#1#2{{{\b{#1}}^{\lower 2.5pt\hbox{$\scriptstyle #2$}}}}
\def\dotexp#1#2{{{#1}^{\lower 2.5pt\hbox{$\scriptstyle #2$}}}}


\def\rt2{\sqrt{2}}

\def\grad{\nabla}



\def\CC{{\cal C}}
\def\CD{{\cal D}}

\def\CH{{\cal H}}

\def\CJ{{\cal J}}

\def\CL{{\cal L}}

\def\CQ{{\cal Q}}


\def\1{{\ds 1}}
\def\R{\hbox{$\bb R$}}

\def\Z{\hbox{$\bb Z$}}


\noblackbox

\def\unit{\relax{\rm 1\kern-.26em I}}
\def\nada{\relax{\rm 0\kern-.30em l}}

\noblackbox
\def\IL{\relax{\rm I\kern-.18em L}}
\def\IH{\relax{\rm I\kern-.18em H}}
\def\IR{\relax{\rm I\kern-.18em R}}
\def\IC{\relax\hbox{$\inbar\kern-.3em{\rm C}$}}
\def\IZ{\relax\ifmmode\mathchoice
{\hbox{\cmss Z\kern-.4em Z}}{\hbox{\cmss Z\kern-.4em Z}} {\lower.9pt\hbox{\cmsss Z\kern-.4em Z}}
{\lower1.2pt\hbox{\cmsss Z\kern-.4em Z}}\else{\cmss Z\kern-.4em Z}\fi}

\def\partialslash{\not{\hbox{\kern-2pt $\partial$}}}

\font\manual=manfnt \def\dbend{\lower3.5pt\hbox{\manual\char127}}

\def\IZ{\relax\ifmmode\mathchoice
{\hbox{\cmss Z\kern-.4em Z}}{\hbox{\cmss Z\kern-.4em Z}} {\lower.9pt\hbox{\cmsss Z\kern-.4em Z}}
{\lower1.2pt\hbox{\cmsss Z\kern-.4em Z}}\else{\cmss Z\kern-.4em Z}\fi}
\def\half{{1\over 2}}

\def\bar{\overline}

\def\rt2{\sqrt{2}}
\def\irt2{{1\over\sqrt{2}}}

\def\slashchar#1{\setbox0=\hbox{$#1$}           
   \dimen0=\wd0                                 
   \setbox1=\hbox{/} \dimen1=\wd1               
   \ifdim\dimen0>\dimen1                        
      \rlap{\hbox to \dimen0{\hfil/\hfil}}      
      #1                                        
   \else                                        
      \rlap{\hbox to \dimen1{\hfil$#1$\hfil}}   
      /                                         
   \fi}


\def\figcaption#1#2{\DefWarn#1\xdef#1{Figure~\noexpand\hyperref{}{figure}%
{\the\figno}{\the\figno}}\writedef{#1\leftbracket Figure\noexpand~\xfig#1}%
\medskip\centerline{{\footnotefont\bf Figure~\hyperdef\hypernoname{figure}{\the\figno}{\the\figno}:}  #2 \wrlabeL{#1=#1}}%
\global\advance\figno by1}


\def\savefig{\expandafter\savefigaux\expandafter{\the\figno}}
\def\savefigaux#1#2#3#4{\DefWarn#2%
 \gdef#2{fig.~\hyperref{}{figure}{#1}{#1}}%
 \writedef{#2\leftbracket fig.\noexpand~\xfig#2}%
 \expandafter\gdef\csname savedfig-\string#2\endcsname{%
   \figinsert\figin{\centerline{#4}}%
   \medskip\centerline{\vbox{\baselineskip12pt
       \advance\hsize by -1truein\noindent\wrlabeL{#2=#2}
       \footnotefont%
       {\bf Fig.~\hyperdef\hypernoname{figure}{#1}{#1}:} #3}}%
   \bigskip\endinsert}%
 \global\advance\figno by1}
\def\putfig#1{\csname savedfig-\string#1\endcsname}
\noblackbox

\input epsf


\def\bb{
\font\tenmsb=msbm10
\font\sevenmsb=msbm7
\font\fivemsb=msbm5
\textfont1=\tenmsb
\scriptfont1=\sevenmsb
\scriptscriptfont1=\fivemsb
}




\def\hat{\widehat}

\def\bar{\overline}
\def\b{\bar}
\def\bsq#1{{{\b{#1}}^{\lower 2.5pt\hbox{$\scriptstyle 2$}}}}
\def\bexp#1#2{{{\b{#1}}^{\lower 2.5pt\hbox{$\scriptstyle #2$}}}}
\def\dotexp#1#2{{{#1}^{\lower 2.5pt\hbox{$\scriptstyle #2$}}}}


\def\rt2{\sqrt{2}}
\def\half {{1 \over 2}}

\def\grad{\nabla}



\def\CC{{\cal C}}
\def\CD{{\cal D}}

\def\CH{{\cal H}}

\def\CJ{{\cal J}}

\def\CL{{\cal L}}

\def\CQ{{\cal Q}}


\def\1{{\ds 1}}
\def\R{\hbox{$\bb R$}}

\def\Z{\hbox{$\bb Z$}}


\noblackbox

\def\unit{\relax{\rm 1\kern-.26em I}}
\def\nada{\relax{\rm 0\kern-.30em l}}



\noblackbox
\def\IL{\relax{\rm I\kern-.18em L}}
\def\IH{\relax{\rm I\kern-.18em H}}
\def\IR{\relax{\rm I\kern-.18em R}}
\def\IC{\relax\hbox{$\inbar\kern-.3em{\rm C}$}}
\def\IZ{\relax\ifmmode\mathchoice
{\hbox{\cmss Z\kern-.4em Z}}{\hbox{\cmss Z\kern-.4em Z}} {\lower.9pt\hbox{\cmsss Z\kern-.4em Z}}
{\lower1.2pt\hbox{\cmsss Z\kern-.4em Z}}\else{\cmss Z\kern-.4em Z}\fi}

\def\CD {{\cal D}}

\def\CJ {{\cal J}}
\def\partialslash{\not{\hbox{\kern-2pt $\partial$}}}

\def\CL {{\cal L}}

\def\CH {{\cal H}}
\def\CC {{\cal C}}


\def\CQ {{\cal Q }}

\font\manual=manfnt \def\dbend{\lower3.5pt\hbox{\manual\char127}}

\def\IZ{\relax\ifmmode\mathchoice
{\hbox{\cmss Z\kern-.4em Z}}{\hbox{\cmss Z\kern-.4em Z}} {\lower.9pt\hbox{\cmsss Z\kern-.4em Z}}
{\lower1.2pt\hbox{\cmsss Z\kern-.4em Z}}\else{\cmss Z\kern-.4em Z}\fi}
\def\half {{1\over 2}}

\def\bar{\overline}

\def\CH{{\cal H}}

\def\rt2{\sqrt{2}}
\def\irt2{{1\over\sqrt{2}}}

\def\hat{\widehat}
\def\slashchar#1{\setbox0=\hbox{$#1$}           
   \dimen0=\wd0                                 
   \setbox1=\hbox{/} \dimen1=\wd1               
   \ifdim\dimen0>\dimen1                        
      \rlap{\hbox to \dimen0{\hfil/\hfil}}      
      #1                                        
   \else                                        
      \rlap{\hbox to \dimen1{\hfil$#1$\hfil}}   
      /                                         
   \fi}


\def\figcaption#1#2{\DefWarn#1\xdef#1{Figure~\noexpand\hyperref{}{figure}%
{\the\figno}{\the\figno}}\writedef{#1\leftbracket Figure\noexpand~\xfig#1}%
\medskip\centerline{{\footnotefont\bf Figure~\hyperdef\hypernoname{figure}{\the\figno}{\the\figno}:}  #2 \wrlabeL{#1=#1}}%
\global\advance\figno by1}


\lref\WilliamsonOFU{
  D.~J.~Williamson, Z.~Bi and M.~Cheng,
  ``Fractonic Matter in Symmetry-Enriched U(1) Gauge Theory,''
[arXiv:1809.10275 [cond-mat.str-el]].
}

\lref\AganagicTVX{
  M.~Aganagic, K.~Costello, J.~McNamara and C.~Vafa,
  ``Topological Chern-Simons/Matter Theories,''
[arXiv:1706.09977 [hep-th]].
}

\lref\PretkoJBI{
  M.~Pretko,
  ``The Fracton Gauge Principle,''
Phys.\ Rev.\ B {\bf 98}, no. 11, 115134 (2018).
[arXiv:1807.11479 [cond-mat.str-el]].
}

\lref\RadzihovskyJDO{
  L.~Radzihovsky and M.~Hermele,
  ``Fractons from vector gauge theory,''
[arXiv:1905.06951 [cond-mat.str-el]].
}

\lref\KogutWT{
  J.~B.~Kogut,
  ``An Introduction to Lattice Gauge Theory and Spin Systems,''
Rev.\ Mod.\ Phys.\  {\bf 51}, 659 (1979)..
}

\lref\GaiottoKFA{
  D.~Gaiotto, A.~Kapustin, N.~Seiberg and B.~Willett,
  ``Generalized Global Symmetries,''
JHEP {\bf 1502}, 172 (2015).
[arXiv:1412.5148 [hep-th]].
}
\lref\BalentsZZ{
  L.~Balents, M.~P.~A.~Fisher and S.~M.~Girvin,
  ``Fractionalization in an easy-axis Kagome antiferromagnet,''
Phys.\ Rev.\ B {\bf 65}, 224412 (2002).
[cond-mat/0110005 [cond-mat.str-el]].
}
\lref\KapustinGUA{
  A.~Kapustin and N.~Seiberg,
  ``Coupling a QFT to a TQFT and Duality,''
JHEP {\bf 1404}, 001 (2014).
[arXiv:1401.0740 [hep-th]].
}

\lref\KitaevWR{
  A.~Y.~Kitaev,
  ``Fault tolerant quantum computation by anyons,''
Annals Phys.\  {\bf 303}, 2 (2003).
[quant-ph/9707021].
}
\lref\NicolisIN{
  A.~Nicolis, R.~Rattazzi and E.~Trincherini,
  ``The Galileon as a local modification of gravity,''
Phys.\ Rev.\ D {\bf 79}, 064036 (2009).
[arXiv:0811.2197 [hep-th]].
}

\lref\QRH{M.~Qi, L.~Radzihovsky, M.~Hermele, ``$p$-string condensation mechanisms for fracton phases from a higher-symmetry viewpoint,''  to appear.}

\lref\NandkishoreSEL{
  R.~M.~Nandkishore and M.~Hermele,
  ``Fractons,''
Ann.\ Rev.\ Condensed Matter Phys.\  {\bf 10}, 295 (2019).
[arXiv:1803.11196 [cond-mat.str-el]].
}

\Title {\vbox{}}
{\vbox{\centerline{Field Theories With a}
\centerline{}
\centerline{Vector Global Symmetry}
\vskip7pt
}}

\centerline{Nathan Seiberg }
\bigskip
\centerline{{\it School of Natural Sciences, Institute for Advanced Study, Princeton, NJ 08540, USA}}

\bigskip
\vskip.1in \vskip.1in
\centerline{\bf Abstract}
\noindent
Motivated by recent discussions of fractons, we explore nonrelativistic field theories with a continuous global symmetry, whose charge is a spatial vector.  We present several such symmetries and demonstrate them in concrete examples.  They differ by the equations their Noether currents satisfy.  Simple cases, other than the translation symmetry, are an ordinary (relativistic) one-form global symmetry and its nonrelativistic generalization.  In the latter case the conserved charge is associated with a codimension-one spatial manifold, but it is not topological.  More general examples involve charges that are integrated over the entire space.  We also discuss the coupling of these systems to gauge fields for these symmetries.  We relate our examples to known continuum and lattice constructions.

\vfill

\Date{September 2019}

\listtoc
\writetoc

\newsec{Introduction and summary}

The goal of this note is to discuss some special global symmetries in nonrelativistic field theories in $D$ spatial dimensions.  Specifically, we will be interested in global symmetries that carry a spatial vector index $i$.  For simplicity, we will limit ourselves to $U(1)$ symmetries and study their Noether currents and conserved charges.\foot{We will see that in some cases the $U(1)$ symmetry is actually noncompact, i.e.\ $\R$.  The transformation parameter is a real number rather than being periodic.} This discussion is generalized easily to $\Z_N$.  The main difference is that in that case there is no Noether current and no conserved charge, but an appropriate exponential of the conserved charge does exist.

This work was motivated by recent exciting ideas about fractons (for a review, see \NandkishoreSEL\ and references therein) and in particular \refs{\PretkoJBI,\WilliamsonOFU}, where similar symmetries have played a crucial role.  Below we will explain the relation between these approaches and ours.

\subsec{Review of ordinary $U(1)$ global symmetry}

Let us start by reviewing the ordinary (zero-form) $U(1)$ symmetry. It is associated with a conserved Noether current.  In relativistic notation we have\foot{Throughout this note we will denote the time coordinate by $x^0$ and the spatial coordinates by $x^i$ with $i=1,...,D$.  We will use Greek letters $\mu,\nu,...$ to combine the indices $0$ and $i$ to denote spacetime indices.  Nowhere in this note will we discuss curved spacetime.  Therefore, when we will raise or lower the time index we will simply multiply by minus one. Except when stated explicitly, we will be interested in flat, noncompact, $D$-dimensional space $\R^D$. We will keep track of upper and lower spatial indices so that it is easy to generalize our discussion to other spatial geometries.}
\eqn\zeroformr{\partial_\mu J^\mu=0}
and in nonrelativistic terms it is
\eqn\Uzc{\partial_0J_0-\partial_iJ^i=0~.}
The conserved charge is an integral over all of space
\eqn\UzQ{Q=\int_{space} J_0~.}
The charged observables are local operators acting at a point.

The form of the current is subject to a known ambiguity, known as ``improvement transformations''
\eqn\Jimprove{\eqalign{
&J_0\to J_0 + \partial_i X^i\cr
&J^j\to J^j + \partial_0 X^j + \partial_i Y^{[ij]}~.}}
with some well defined operators $X^i$, $Y^{[ij]}$, where $[ij]$ means that the indices are antisymmetric. Since $J_0$ is shifted by a total spatial derivative, the charge \UzQ\ is not modified.  (This assumes sufficient fall off at infinity.)  The shift of $J^j$ is such that the current remains conserved.

A conserved current $J^\mu$ can be coupled to a gauge field $A_\mu$.  The minimal coupling is
\eqn\minc{A_\mu J^\mu~.}
Nonminimal couplings include other gauge invariant terms linear in $A_\mu$.  Specifically, terms of the form
\eqn\nonmi{F_{\mu\nu}Z^{[\mu\nu]}}
with $F_{\mu\nu}=\partial_\mu A_\nu - \partial_\nu A_\mu$ are related to the improvement transformation \Jimprove.

We also note that depending on the details of the theory the minimal coupling \minc\ is often accompanied by ``seagull terms'' that are higher order in $A_\mu$.

\subsec{A symmetry with a vector index and a summary of this note}

If the charge has a vector index, the conserved Noether current must have two indices.  A well known example of such a charge is the momentum $P^i=\int_{space}T_0^i$.  Other standard examples are based on a conserved current with two antisymmetric indices
\eqn\relative{\partial_\mu J^{[\mu\nu]}=0~.}
Here we will follow \refs{\KapustinGUA\GaiottoKFA} and interpret it as a one-form global symmetry.  In nonrelativistic notation \relative\ is
\eqn\Uonec{\eqalign{
&\partial_0J_0^j-\partial_iJ^{[ij]}=0\cr
&\partial_i J_0^i=0~.}}
From a nonrelativistic perspective the first equation states the conservation and the second equation is an additional condition.

The conserved charges are associated with codimension-one manifolds in space $\CC$
\eqn\UonecQ{\CQ(\CC)=\int_\CC J_0^j n_j}
with $n_j$ a unit vector normal to $\CC$.  The second condition in \Uonec\ guarantees that these charges do not change under small changes of $\CC$ .  

In section 2, we will discuss these charges in detail and will stress that the charged operators are associated with lines.  We will demonstrate the symmetry in continuum $U(1)$ gauge theory and in its lattice version.  Here the second condition in \Uonec\ is Gauss law and the conserved charge \UonecQ\ is the electric flux through $\CC$.

In sections 3 and 4, we will present generalizations of the relativistic one-form symmetry \Uonec\ in a hierarchical way.

In sections 3, we will generalize \Uonec\ to a current that satisfies the conservation equation (the first condition in \Uonec), but does not satisfy the second condition in \Uonec
\eqn\Uonecn{\eqalign{
&\partial_0J_0^j-\partial_iJ^{[ij]}=0\cr
&G\equiv \partial_i J_0^i\not=0~.}}  
This is inconsistent with relativistic invariance, but is consistent in a nonrelativistic system.  One important fact, which follows trivially from the defining equation \Uonecn\ is that $G$ is conserved at every point
\eqn\Gcons{\partial_0 G=0~.}
Unlike the more special case \Uonec, where $G$ vanishes, here it is nonzero, but it is conserved.

The conserved charges are still given by \UonecQ, but since $G$ no longer vanishes, they change under small changes of $\CC$.

The conserved global symmetry \Gcons\ is quite unusual.  There is a separate conservation at every point in space.  Equivalently, we have a conserved current with charge
\eqn\Gcharge{Q_{C(x^i)}=\int_{space}C(x^i)G}
for any space-dependent function $C(x^i)$.

In section 3, we will also present continuum and lattice examples of systems with that symmetry.  They are obtained by coupling a $U(1)$ gauge theory to special matter fields.  In this interpretation $G$ in \Uonecn\ is the divergence of the electric field, i.e.\ Gauss law.  In the system without matter we have \Uonec\ and $G$ vanishes.  Now, it is nonzero because of the presence of charged matter.

This matter system is quite peculiar, because it leads to the conserved current \Uonecn. It can be thought of as infinitely massive charged scalar fields.  They are pinned and are immobile.  Yet, this matter system is not merely a product of independent systems at different points in space.

In section 4, we will generalize the symmetry further by considering $J^{ij}$ in \Uonecn\ whose indices are not necessarily antisymmetric.  We will impose
\eqn\Uvecc{\partial_0J_0^j-\partial_iJ^{ij}=0~.}
Unlike \Uonec\ where $G=\partial_iJ_0^i$ vanishes and unlike the more general case \Uonecn\ where $G$ is nonzero but conserved, here it is not even conserved.

Since $G$ is not conserved, the charges associated with codimension-one manifolds $\CC$ \UonecQ\ are also not conserved.  Instead, the only conserved charge is an integral over the entire space
\eqn\veccon{Q^j=\int_{space}J_0^j ~.}
We will show how the special matter system in section 3 can be modified to account for this change in the current.

We will also consider the coupling of the systems with the global symmetry \Uonec \Uonecn \Uvecc\ to appropriate gauge fields.  The minimal coupling is similar to \minc\ and ``nonminimal'' terms analogous to \nonmi\ are associated with the freedom to perform ``improvement transformations'' of the current (similar to \Jimprove)
\eqn\improcc{\eqalign{
&  J_0^j\to J_0^j+ \partial_k X^{kj} \cr
&  J^{ij}\to J^{ij} + \partial_0 X^{ij} +\partial_k Y^{[ki]j}~.}}

Finally, in an appendix we will consider these peculiar matter systems on their own (without the $U(1)$n gauge field) and show that their canonical quantization leads to a strange Hilbert space with an infinite vacuum degeneracy!

\newsec{Relativistic one-form global symmetry}

\subsec{The symmetry}  

There has been a lot of work on conserved currents with antisymmetric indices.  Here we follow \GaiottoKFA\ and interpret it as a one-form global symmetry.  

We start by describing the relativistic symmetry in a $D+1$-dimensional spacetime.  The conservation equation \relative
\eqn\relativea{\partial_\mu J^{[\mu\nu]}=0~}
means that an observable obtained by integrating the current over a codimension-two manifold $\CC$ in spacetime
\eqn\chargere{\CQ(\CC)=\int J^{[\mu\nu]}n_{[\mu\nu]}}
with $n_{[\mu\nu]}$ a unit vector normal to $\CC$ has special properties.  It does not change under small deformations of $\CC$.  We refer to this property of the operator as being topological.  The only way its correlation functions can change under deformations of $\CC$ is when the deformation crosses another operator.

Given our $D+1$-dimensional spacetime, the only meaningful crossing of $\CC$ with another operator is when this operator is a line operator.  (Below, when we demonstrate this symmetry in a concrete field theory this line operator will be a Wilson line.)

Let us repeat this discussion in nonrelativistic terms.  The conservation equation \relativea\ becomes \Uonec
\eqn\Uoneca{\eqalign{
&\partial_0J_0^j-\partial_iJ^{[ij]}=0\cr
&G=\partial_i J_0^i=0~.}}
The first equation is a conservation equation.  The second equation, which is related to the first one in the relativistic theory \relativea\ might seem unnatural.  (Indeed, below we will relax it.)  

In the nonrelativistic theory, operators act at a given time, but they might be integrated along some region in space.\foot{Insertions that are integrated along the time direction should be interpreted as defects rather than operators acting on the Hilbert space.}  Therefore, an operator \chargere\ at fixed time is \UonecQ
\eqn\UonecQa{\CQ(\CC)=\int_\CC J_0^j n_j}
with $\CC$ a codimension-one manifold in space and $n_j$ is orthogonal to it.  If $\CC$ encloses a compact region without any operator insertion,
\eqn\UonecQas{\CQ(\CC)=\int_\CC J_0^j n_j=\int_{\hat\CC}\partial_j J_0^j =0 ~,}
where $\CC=\partial\hat\CC$ is the boundary of $\hat \CC$ and we used the second equation in \Uoneca.  The only way to get a nonzero answer is when there is no such $\hat\CC$, i.e.\ when the underlying space has nontrivial topology, or when $\CC$ is noncompact.  Even on flat $\R^D$, there can still be nontrivial topology if the theory includes ``defects.''  These are line insertions along the time direction.  They act as delta function sources violating $G=\partial_jJ_0^j=0$.  Then, $\CQ(\CC)$ receives contributions only from these points.  In the example below, these insertions will be classical background charges.

It is standard to gauge this global symmetry by coupling it to a two-form gauge field $B_{\mu\nu}$.  The minimal coupling, similar to \minc, is 
\eqn\Bmunuc{B_{\mu\nu}J^{[\mu\nu]}~.}
The conservation equation \relativea\ guarantees that it is gauge invariant under
\eqn\Bmunug{B_{\mu\nu} \to B_{\mu\nu}+\partial_\mu c_\nu - \partial_\nu c_\mu ~.}
It is important that $c_\mu$ is a $U(1)$ gauge field.  One way to see that is that $B_{\mu\nu}$ does not transform under a pure gauge $c_\mu=\partial_\mu \lambda$.  This also means that if spacetime is more complicated, there are gauge transformations of $B_{\mu\nu}$ with nonzero ``flux'' of $c_\mu$.
The gauge invariant field strength of $B_{\mu\nu}$ is $H_{\mu\nu\rho}=\partial_{[\mu} B_{\nu\rho]}$ and using that we can add to the Lagrangian terms of the form $H_{\mu\nu\rho}^2$ and other couplings.

This discussion of gauging the global symmetry is essentially unchanged in a nonrelativistic system.  The only difference is that the coefficients of $H_{0ij}^2$ and $H_{ijk}^2$ in the Lagrangian do not have to be the same.

\subsec{Example: $U(1)$ continuum gauge theory}

The simplest example of a system with a $U(1)$ one-form global symmetry is an ordinary $U(1)$ gauge theory with the continuum Lagrangian\foot{In a nonrelativistic system, the term with the electric field and the term with the magnetic field can have different coefficients, but in order to simplify the expressions, we are going to take these coefficients to be equal.  It is trivial to extend the discussion to the more generic case.}
\eqn\Uoneco{-{1\over 4g^2} F_{\mu\nu}F^{\mu\nu}~.}
The equation of motion of $A_\mu$
\eqn\Amueom{\partial_\mu F^{\mu\nu}=0}
means that we have the conservation \relativea\ with the identification
\eqn\FJid{J^{[\mu\nu]}={1\over g^2}F^{\mu\nu}~.}
In nonrelativistic terms, the equation of motion of $A_i$ leads to the first equation in \Uoneca\ and the equation of motion of $A_0$ leads to the second equation in \Uoneca, which is now interpreted as Gauss law.  (Hence the notation $G$.)

The conserved charge \UonecQa\ is the electric flux through $\CC$.  Clearly, it does not change under deformations of $\CC$,unless we cross a background charge.  This crossing of a background charge is the manifestation of the general discussion following \UonecQas.  It is clear that the Wilson lines $e^{i\oint A}$ are charged under the one-form symmetry.  The normalization of the current \FJid\ is set such that its charge is one.

This system also has a magnetic $D-2$-form global symmetry \GaiottoKFA, but we will ignore it in this discussion.

Let us consider the effect of the charge operator $\CQ(\CC)$ on the gauge fields in more detail.  The group element generated by $\CQ(\CC)$ is $e^{i\beta\CQ(\CC)}$ and the charged operators are Wilson lines $W_q(\ell)=e^{iq\int_\ell A}$ with $\ell$ a line in space.  Then the group action is
\eqn\Uonegroup{e^{i\beta\CQ(\CC)} W_q(\ell) = W_q(\ell)e^{i\beta\CQ(\CC)} e^{iq\beta I(\ell,\CC)}}
with $I(\ell,\CC)$ the oriented number of times $\ell$ pierces $\CC$.  This can be interpreted to mean that $\CQ(\CC)$ shifts the gauge field by a flat gauge field \GaiottoKFA.

The coupling to a two-form gauge field \Bmunuc\ is standard.  The gauge field $A_\mu$ transforms under \Bmunug\ as (see the discussion around \Uonegroup)
\eqn\Amutr{A_\mu \to A_\mu  - c_\mu}
and then we replace \Uoneco\ by
\eqn\Uonecog{-{1\over 4 g^2} (F_{\mu\nu}+B_{\mu\nu})(F^{\mu\nu}+B^{\mu\nu}) - {1\over g_H^2} H_{\mu\nu\rho}H^{\mu\nu\rho}~.}
This includes a $B_{\mu\nu}^2$ seagull term we mentioned above.  (As for $F_{\mu\nu}F^{\mu\nu}$, there can be different coefficients for the kinetic terms that are not related by the spatial Euclidean symmetry.)

\subsec{Example: $U(1)$ lattice gauge theory}

Let us repeat this discussion on the lattice.\foot{A classic review of lattice gauge theory is \KogutWT.}

In a Hamiltonian formalism the theory is described by group elements $U_l=e^{iA_l}$ and their conjugate electric field operators $E_l$ on the links, which are labeled by $l$.  The Hamiltonian includes a sum over plaquette terms
\eqn\Updef{ U_p=e^{i\sum_{l\in \{l_p\}} \epsilon_l A_l}=e^{i\nabla\times A}}
(and their complex conjugates), where the sum in the exponent is an oriented sum (with $\epsilon_l=\pm 1$) over the links $\{l_p\}$ around the plaquette $p$.  The kinetic terms in the Hamiltonian are a sum over  $E_l^2$.    In addition, we need to impose the Gauss law constraint
\eqn\Gausslaw{ G=\sum_{l\in \{l_s\}}  \epsilon_l E_l = \nabla\cdot E=0~,}
where the sum is an oriented sum (with $\epsilon_l=\pm1$) over the links $\{l_s\}$ emanating from the site $s$.

More generally, we can introduce probe electric charges $q_i$ at the sites $s_i$ by modifying \Gausslaw
\eqn\Gausslawm{\sum_{l\in \{l_s\}}  \epsilon_l E_l = \nabla\cdot E=\sum_i q_i \delta_{s,s_i}~.}
We emphasize that these are probe charges, rather than dynamical particles.  (This can be generalized to probe charges whose positions change in time.)

This system has an electric one-form global symmetry \GaiottoKFA\ analogous to the continuum current \relativea.  And the conserved charge \UonecQa\ $\CQ(\CC)$ is
\eqn\latticeone{\CQ(\CC) = \sum_{l\in \CC}\epsilon_l E_l ~,}
where $l$ labels the links pierced by $\CC$ and again $\epsilon_l=\pm1$.  Clearly, $\CQ(\CC)$ commutes with the kinetic term, which depends only on $E_l$.  To see that it commutes with the plaquette terms, focus on a given plaquette $p$.  $\CC$ pierces an even number of the links around it.  Then the commutator of \latticeone\ with $U_p$ receives an even number of contributions and they cancel each other.  Therefore, $\CQ(\CC)$ commutes with the Hamiltonian.

A key fact about $\CQ(\CC)$ is that it is not only conserved, but it is also topological.  Small changes in $\CC$ do not change the correlation functions, provided they do not cross any insertion of an operator.  This fact follows from Gauss law \Gausslaw.  All this is as in the discussion around  \UonecQas\ and is identical to the continuum discussion above.

More generally, if probe particles are present, \Gausslawm\ means that $\CQ(\CC)$ changes when it crosses a site with nonzero charge.  Equivalently, the operator $\CQ(\CC)$ measures the total charge enclosed by $\CC$, i.e.\ the sum of the charges $q_i$ in \Gausslawm\ in that region.  This is consistent with the fact that $\CQ(\CC)$ measures the electric flux through $\CC$.

The existence of this electric one-form global symmetry is tied to the fact that the system does not include charged dynamical degrees of freedom.  When such matter fields are present, the charges $q_i$ in \Gausslawm\ are dynamical,  they fluctuate, and thus ruin the symmetry.

When the gauge group is compact,\foot{Here we use the standard mathematical terminology, which is also used in high energy physics.  The meaning of a compact gauge group in condensed matter physics is different.} the analogous continuum theory also has a magnetic $D-2$-form global symmetry, which is generated by integrating the magnetic field along a surface $\Sigma$, $\int_\Sigma F$.  The lattice system typically does not have that symmetry.

\newsec{Nonrelativistic one-form global symmetry}

\subsec{The symmetry}

In this section we consider the nonrelativistic symmetry \Uonecn
\eqn\Uonecna{\partial_0J_0^j-\partial_iJ^{[ij]}=0}
without imposing the second condition $G\equiv \partial_i J_0^i$.  Clearly, this is not possible in a relativistic system, but it is perfectly natural in a nonrelativistic system.

Applying $\partial_j$ to the conservation equation and using the antisymmetry of the spatial indices we derive
\eqn\Gconsa{\partial_0 G=0~.}
Therefore, $G$ does not vanish, but it is conserved at every point. 

This is an unusual conservation law and many of the peculiarities of this symmetry stem from it.  In particular, as we said around \Gcharge,
\eqn\Gcharges{Q_{C(x^i)}=\int_{space}C(x^i)G}
is conserved for any space-dependent function $C(x^i)$.

Instead of thinking about the charges \Gcharges, we can follow the discussion of the relativistic system.  There we considered conserved charges that are given by integrals over a codimension-one spatial manifold $\CC$
\eqn\UonecQaa{\CQ(\CC)=\int_\CC J_0^j n_j}
with $n_j$ a unit vector normal to $\CC$.\foot{Actually, when $\CC$ is a nontrivial cycle in space, \UonecQaa\ provides additional information beyond \Gcharges.}  In the relativistic case we used the fact that $G=\partial_jJ_0^j=0$ to prove that this expression is topological.  Now, this is no longer true.  $\CQ(\CC)$ changes under small deformations of $\CC$ even if they do not cross defects.

In a sense, this symmetry is a lot larger than the relativistic one. We can say that the relativistic symmetry is a quotient of the nonrelativistic symmetry by $G=0$.  The quotient identifies $\CQ(\CC)$ with $\CC$ in the same homotopy class.\foot{The authors of \QRH\ refer to the relativistic symmetry as a nonfaithfully acting version of this nonrelativistic symmetry.}

In flat space, there is an interesting subgroup generated by a cruder charge
\eqn\cruder{Q^j=\int_{space}J_0^j~.}
It is an integral of $\CQ(\CC)$ over all codimension-one manifolds $\CC$ that are normal to the $j$ direction.  We can also think of $Q^j$ as $Q_{C(x^i)}$ of \Gcharges\ with $C(x^i)=x^i$.  Exponentiating $Q^j$ leads to a symmetry group element
\eqn\crudere{e^{ic_jQ^j}~}
with constant $c_j$.  As we will see below, these constants are arbitrary real numbers and they are not subject to any identification.

Another consequence of the larger symmetry group is that unlike the relativistic symmetry, whose charged objects are lines, here point operators can also transform under the symmetry.  This is clear because the local operator $G$ is nontrivial.  Below we will see examples of this.

Next, we would like to discuss gauging of this symmetry.  As in the relativistic case we introduce gauge fields $B_{\mu\nu} $ with minimal coupling \Bmunuc
\eqn\Bmunucn{2B_{0j}J_0^j-B_{ij}J^{[ij]}~.}
However, now the conservation equation \Uonecna\ guarantees gauge invariance under
\eqn\Bmunugn{\eqalign{
&B_{0j} \to B_{0j}+\partial_0 c_j  \cr
&B_{ij} \to B_{ij}+\partial_i c_j - \partial_j c_i ~}}
but without a shift by $c_0$ in \Bmunug.  

The lack of gauge transformation with $c_0$ originates from the fact that we did not impose $G=0$ and it is closely related to the fact that the charges $\CQ(\CC)$ are not topological.  Another consequence of that is that unlike the relativistic discussion around \Bmunug, here the gauge parameter $c_i$ is not a $U(1)$ gauge field.  Indeed, $c_i=\partial_i\lambda$ leaves $B_{ij}$ invariant, but it shifts $B_{0j}$ by $\partial_0\partial_j\lambda$.  As a result, $c_i$ must be single valued.
This fact is consistent with the comment after \crudere\ that in the system before gauging, $c_i$ are arbitrary real numbers and are not subject to any identification.

Finally, as in the relativistic theory, we can use the gauge invariant field strength $H_{\mu\nu\rho}$ to add a kinetic terms for $B_{\mu\nu}$.

\subsec{Example: $U(1)$ lattice gauge theory with energetically imposed Gauss law}

It is common to follow \refs{\KitaevWR,\BalentsZZ} and change the standard lattice gauge theory by relaxing the Gauss law constraint \Gausslaw\ and replacing it by a term that imposes it energetically, e.g.\ adding to the Hamiltonian a sum over all the sites $s$
\eqn\divEs{\sum_s (\nabla\cdot E)^2~.}  
This means that the Hilbert space includes dynamical high energy excitations carrying electric charge.  The presence of these massive charges affects the one-form symmetry generated by \latticeone.

When Gauss law is not imposed, $ \nabla\cdot E$ is nonzero.  However, it is easy to check that it, and more generally $\CQ(\CC)$, are still conserved. 

Unlike the case of $\nabla\cdot E=0$, now there are local physical operators that are charged under $ \nabla\cdot E$ and $\CQ(\CC)$ .  For example, the link element $U_l$ is charged under $ \nabla\cdot E$ at the two ends of $l$.  When $\nabla\cdot E=0$ was imposed, these operators were not gauge invariant.  Now, they are consistent physical operators.  We can interpret it to mean that the system is a standard gauge theory with massive charged particles and this operator describes a pair of charged particles at its two ends such that it is gauge invariant.

Let us summarize.  The standard pure gauge system has a one-form symmetry \relative. The system with the energetically imposed Gauss law has the more generic, nonrelativistic one-form symmetry \Uonec.  Explicitly, $G=\partial_iJ_0^i$ is proportional to Gauss law.  In the standard gauge theory it vanishes, as in \relative.  And when it is imposed energetically, it is conserved, but does not vanish, as in \Uonec.

It is important to stress that the $U(1)$ lattice gauge theory with energetically imposed Gauss law is not a generic $U(1)$ gauge theory with massive charged particles.  The generic theory does not have any global symmetry with a vector charge, but the theory with the energetically imposed Gauss law does have such a symmetry.

This interpretation is consistent with the analysis in \WilliamsonOFU.  That paper focused on the global symmetry generated by the sum of the electric field over the entire lattice, which is the lattice version of the cruder charge \cruder. (The more detailed symmetry obtained by summing the electric field on a codimension-one subspace is mentioned in \WilliamsonOFU.)  The main difference between our analysis and \WilliamsonOFU\ is in the way this symmetry is gauged.  Our gauging used the gauge fields $B_{\mu\nu}$.  The gauging in \WilliamsonOFU\ also involves fields on the plaquettes, but they are symmetric (rather than antisymmetric) and there are also additional fields.

\subsec{Example: coupling a continuum $U(1)$ gauge theory to a special matter system}

As we said above, the lattice gauge theory with Gauss law imposed energetically can be thought of as an ordinary lattice gauge theory, but with massive charged matter.  However, in order to preserve the nonrelativistic global symmetry, the charged matter should be special.

Let us look for a continuum field theory for that system.  We would like to start with the standard continuum $U(1)$ gauge theory (section 2.2) and add a matter theory to it, such that it respects the nonrelativistic one-form symmetry.  We want to retain \FJid\
\eqn\FJids{J^{[\mu\nu]}={1\over g^2}F^{\mu\nu}~}
i.e.\ impose \Uonecna
\eqn\Uonecnaa{\partial_0J_0^j-\partial_iJ^{[ij]}={1\over g^2}(\partial_0F_0^{~j}-\partial_iF^{[ij]} )=0~,}
but relax Gauss law 
\eqn\nogauss{G=\partial_iJ_0^i={1\over g^2}\partial_i F_0^{~i} \not=0~.}

This means that the $U(1)$ gauge theory is coupled to a matter theory with a global $U(1)$ symmetry with a current $\CJ_\mu$
\eqn\mattercond{\CJ_0\not=0 \qquad, \qquad \CJ^i=0 \qquad,\qquad \partial_0\CJ_0=0~.}
(Note that if $\CJ^i=\partial_j \CJ^{[ji]}\not=0$, the current is still conserved, but it can be improved \Jimprove\ to this current.)

Such a matter system must be peculiar.  The fact that it has a conserved nonzero operator $\CJ_0$ means that not only $\CJ_0$ is time independent, but $C(x^i)\CJ_0$ is time independent for every function of the spatial coordinates $C(x^i)$.

Denoting the matter theory Lagrangian by $\CL_0$ we use minimal coupling \minc\ and the $U(1)$ Lagrangian \Uoneco\ to write
\eqn\Lononr{\CL_1=\CL_0- {1\over 4 g^2} F_{\mu\nu}F^{\mu\nu} + A_0\CJ_0~.}
As above, for simplicity we took the coefficient of $E^2$ and of $B^2$ to be the same and we neglected higher order terms and possible seagull terms.  Also, we could add to the Lagrangian $ A_0\rho_0$ with $\rho_0$ a classical, time-independent, background charge density, but in order to simplify the equations, we did not do it.

The equations of motion of $A_j$ derived from \Lononr\ lead to \Uonecnaa.  Therefore, this theory has the nonrelativistic one-form symmetry and the entire discussion in section 3.1 applies to it.

It is easy to construct an explicit example of a matter theory with these properties.  The charged matter is a massive complex scalar field $\Phi$.  The global $U(1)$ symmetry acts on it as $\Phi\to e^{i\alpha} \Phi$ and therefore, the symmetry with arbitrary $C(x^i)$ acts as
\eqn\Cxiaction{\Phi \to e^{iC(x^i)}\Phi~.}

It is standard in nonrelativistic field theory to remove a term of the form $m^2|\Phi|^2$ using a redefinition of $\Phi$ by a time dependent phase and then rescale $\Phi$.  Then, the leading order terms are
\eqn\Lzeron{
\CL_0=i\bar\Phi \partial_0\Phi - s\partial_i(\bar\Phi\Phi)\partial^i(\bar \Phi\Phi) -\lambda|\Phi|^4  +\cdots ~.}
$s$ and $\lambda$ are real coupling constants and the ellipses represent terms with more derivatives or higher powers of the field.  The most important point about \Lzeron\ is that it does not include a term of the form $\partial_i\Phi\partial^i\bar\Phi$.  Such a term violates \Cxiaction.\foot{C.~Vafa suggested that this Lagrangian is similar to the theory in \AganagicTVX.}

We recognize that 
\eqn\CJex{\CJ_0=\bar\Phi\Phi+\cdots~.} 
 
The combined matter system $\CL_0$ of \Lzeron\ and the $U(1)$ Lagrangian \Uoneco\ as in \Lononr\
\eqn\Lononre{\CL_1=i\bar\Phi (\partial_0-iA_0)\Phi - s\partial_i(\bar\Phi\Phi)\partial^i(\bar \Phi\Phi) -\lambda|\Phi|^4 - {1\over 4 g^2} F_{\mu\nu}F^{\mu\nu} +\cdots~.}
leads to a system with the desired properties.
It is a continuum version of the energetically imposed Gauss law lattice system and provides an explicit continuum realization of the nonrelativistic one-form global symmetry.

The equation of motion of $A_0$ leads to
\eqn\Gex{G={1\over g^2} \partial_iF_0^{~i}= -\CJ_0 = -|\Phi|^2+\cdots~.}

Using this explicit realization of the symmetry it is easy to find gauge invariant operators that transform under it.  Specifically,
\eqn\gaugeinvtr{\CD_i(x)=\bar\Phi(\partial_i - iA_i)\Phi(x) }
transforms nontrivially under $G(x)$ \Gex\ and
\eqn\gaugeinvrts{W(x,y)=\bar\Phi(x)e^{-i\int_x^y A} \Phi(y)}
transforms nontrivially under $G(x)$ and $G(y) $.
Similarly, 
\eqn\cruderee{e^{ic_jQ^j}~}
maps
\eqn\transfoc{\eqalign{
\CD_i=\bar\Phi(\partial_i - iA_i) \Phi \quad &\to \quad \CD_i+ ic_i\bar\Phi\Phi \cr W(x,y)=\bar\Phi(x)e^{-i\int_x^y A} \Phi(y)\quad &\to \quad e^{ic_i(y-x)^i}W(x,y)
~.}}
This demonstrates the point we made after \crudere\ that the coefficients $c_i$ are not subject to any identification.

We emphasize again that the lack of a standard $\partial_i\Phi\partial^i\bar\Phi$ term means that the system is quite nonstandard.  In appendix A, we will exhibit some of its peculiarities that originate from this fact.

\newsec{A more general vector symmetry}

\subsec{The symmetry}

In the relativistic and the nonrelativistic symmetries we discussed above, the indices in $J^{ij}$ were antisymmetric.  In this section we consider a more general case \Uvecc\
\eqn\Uvecc{\partial_0J_0^j-\partial_iJ^{ij}=0~,}
with no restriction on the spatial indices. $J^{ij}$ includes an antisymmetric tensor, a scalar, and a traceless symmetric tensor.  (We could further separate the discussion depending on whether the scalar, the traceless symmetric tensor, or both of them are present.)

Since $J^{ij}$ is not antisymmetric, $G=\partial_iJ_0^i$ is not conserved 
\eqn\Gdot{\partial_0G=\partial_i\partial_j J^{ij}\not=0~.}
Related to that, the codimension-one observables $\CQ(\CC)$ are not conserved.  The only conserved charge is the cruder charge obtained as an integral over the entire space \veccon
\eqn\veccon{Q^j=\int_{space} J_0^j ~.}

Comments:
\item{1.} $\Big(I_0=G,~I^j=\partial_iJ^{ij}\Big)$ seems like an ordinary conserved current \Uzc.  However, since $G=\partial_iJ_0^i$ is a total spatial derivative, the charge of this current is trivial.  It can be improved \Jimprove\ to zero.
\item{2.} One might try to generalize the relativistic symmetry \relative\ in a similar way by considering a conserved current $J^{\mu\nu}=J^{[\mu\nu]} +\delta^{\mu\nu}J$ with a scalar $J$.  However, this leads to $\partial_\mu\partial^\mu J=0$ and hence $J$ is a free field.

\bigskip
Next, as in the relativistic discussion following \Bmunuc\ and in the nonrelativistic case following \Bmunucn, we would like to gauge the symmetry.  Unlike the previous cases, here it is not enough to introduce the gauge fields $B_{\mu\nu}$ because the current operator $\Big(J_0^j,\ J^{ij}\Big)$ includes more tensors.  Minimal coupling in this case must be of the form
\eqn\Bmunucnv{2B_{0j}J_0^j-(B_{ij}+S_{ij})J^{ij}}
with an antisymmetric tensor gauge field $B_{ij}$ and a symmetric tensor gauge field $S_{ij}$.
In this case the conservation equation \Uvecc\ guarantees gauge invariance under
\eqn\Bmunugnv{\eqalign{
&B_{0j} \to B_{0j}+\partial_0 c_j  \cr
&B_{ij} \to B_{ij}+\partial_i c_j - \partial_j c_i \cr
&S_{ij} \to S_{ij}+\partial_i c_j +\partial_j c_i~.}}
If either the traceless symmetric component of $J^{ij}$ or the trace vanish, then only the trace of $S_{ij}$ or its traceless part should be introduced.  But otherwise, all the components are needed.  As in the nonrelativistic case above, $c_i$ is not a $U(1)$ gauge field.  

Of course, as in all the previous cases, we can add gauge invariant kinetic terms for these gauge fields and add higher order couplings of them to the original system.  

It will be interesting to explore these gauge fields in detail.

\subsec{Example: coupling a continuum $U(1)$ gauge theory to a special matter system}

Here we will extend the discussion in section 3.3 to find a class of continuum field theories exhibiting the more general vector symmetry \Uvecc.  As in section 3.3, we couple a matter sector with a global $U(1)$ symmetry to the $U(1)$ gauge theory in such a way that the resulting theory has the symmetry \Uvecc.  

We will first discuss the matter theory in general and then give a concrete example based on a charged scalar field.

\bigskip\centerline{\it The matter system}\bigskip

We assume that the matter theory has operators $\CJ _0$ and a symmetric tensor $\CJ^{ij}$ satisfying
\eqn\mattre{\partial_0\CJ _0=\partial_i\partial_j \CJ^{ji}~.}
This generalizes \mattercond\ to nonzero $\CJ^{ij}$.  

Before we couple this matter system to the $U(1)$ gauge theory, we would like to discuss its global symmetries.  First, it has an ordinary global symmetry \Uzc\ 
\eqn\ordinma{J_0=\CJ_0\qquad, \qquad J^i=\partial_j\CJ^{ji}}
with the conserved charge
\eqn\ordinamac{Q=\int _{space} \CJ_0~.}
This is the symmetry that we will soon gauge.

It also has a vector global symmetry \Uvecc\ with currents
\eqn\vectorm{\eqalign{
&J_0^j=x^j\CJ _0 \cr
& J^{ij}=x^j\partial_k\CJ^{ki} - \CJ^{ij}~.}}
It is easy to check, using \mattre, that this current is conserved as in \Uvecc.  The corresponding charge is
\eqn\chargemad{Q^j=\int_{space}J_0^j=\int _{space}x^j\CJ_0~.}
Since $\CJ_0$ is the charge density of the ordinary global $U(1)$ symmetry, \chargemad\ leads us to interpret the global symmetry of \vectorm\chargemad\ as a dipole symmetry.

Clearly, a shift of $\CJ^{ij}$ by an antisymmetric tensor does not affect \mattre\ and the currents \ordinma\ and \vectorm\ change only by an improvement transformation.  Therefore, we can limit our discussion to symmetric $\CJ^{ij}$.

In the matter system in section 3.3, we had a large global symmetry with charge density $C(x^i)\CJ_0$.  Here the symmetry is smaller and only a linear function $C(x^i)=\alpha + c_ix^i$ with constant $\alpha$ and $c_i$ leads to a conserved charge. These transformations are implemented by
\eqn\opssym{U(\alpha) = e^{i\alpha Q} \qquad, \qquad V(c_j) =e^{ic_jQ^j} ~}
with the charges \ordinamac\chargemad.
The explicit $x^i$ dependence of the current \vectorm\  means that the charge $Q^j$ does not commute with spatial translations. Denoting the translation transformation by $P_i$ we have
\eqn\PVP{V(c_i) e^{ir^iP_i}= e^{ir^iP_i}V(c_i)U(c_ir^i)~.}

\bigskip\centerline{\it Gauging the global $U(1)$ symmetry}\bigskip

Next, following \minc, we gauge the ordinary global $U(1)$ symmetry  \ordinma\ordinamac\ by writing the couplings\foot{Pretko's discussion \PretkoJBI\ can be phrased in terms of adding nonstandard gauge fields that couple linearly to $\CJ_0$ and $\CJ^{ij}$ as $A_0\CJ_0 +\half A_{ij}\CJ^{ij}$.  Our coupling here can be recast in this form with $A_{ij}=\partial_iA_j+\partial_jA_i$.  However, not every symmetric $A_{ij}$ can be expressed in terms of a standard gauge field $A_i$.}
\eqn\gaugeLag{
\CL_1(A)= \CL_0 -{1\over  4 g^2} F_{\mu\nu}F^{\mu\nu}  + A_0\CJ_0 -A_i\partial_j\CJ^{ij}~.}
As in section 3, $\CL_0$ is the Lagrangian of the matter theory and $F_{\mu\nu}=\partial_\mu A_\nu-\partial_\nu A_\mu$ is the standard field strength.  Again, for simplicity we set the coefficients of $E^2$ to be the same as the coefficient of $B^2$, we suppressed gauge invariant terms with more derivatives, gauge invariant nonminimal couplings of $F_{\mu\nu}$ to other matter operators, and possible seagull terms.  Also, we did not add to the Lagrangian a coupling $ A_0\rho_0$ to background charge density.

The equations of motion of the gauge fields are
\eqn\fieldseomg{\eqalign{
&{1\over g^2}\partial_j F_{0}^{~j} +\CJ_0=0\cr
&{1\over g^2} \partial_0 F_{0}^{~i} -{1\over g^2} \partial_jF^{ji}+\partial_j\CJ^{ji}=0 ~.
}}

After the gauging, the current of the global symmetry of \ordinma\ordinamac\ becomes trivial.\foot{
More precisely, it can be improved to a trivial current
\eqn\currentims{\eqalign{
&J_0\to J_0 +{1\over g^2}\partial_j F_{0}^{~j}=0\cr
& J^i\to J^i +{1\over g^2}\partial_0 F_{0}^{~i} -{1\over g^2}\partial_j F^{ji}=0~,}}
where we used the equations of motion \fieldseomg.}
The Noether current of the remaining global symmetry can now be written as\foot{
This current is related to \vectorm\ by an improvement transformation.  Use $X^{kj}=-{1\over g^2}x^j F_0^{\ k}$ and $Y^{[ki]j}={1\over g^2} x^j F^{ki }$ in \improcc\ and the equations of motion \fieldseomg\ to find
\eqn\Noethercgi{\eqalign{
&J_0^j= {1\over g^2} F_0^{\ j} \to {1\over g^2} F_0^{\ j} -{1\over g^2}\partial _k\left(x^j F_0^{\ k}\right)=  x^j \CJ_0\cr
&J^{ij}= {1\over g^2 }F^{ij} -\CJ^{ji}\to {1\over g^2 }F^{ij} - \CJ^{ij}-{1\over g^2}\partial_0\left(x^j F_{0}^{~ i}\right)+ {1\over g^2}\partial_k\left( x^jF^{ki }\right)
=x^j\partial_k\CJ^{ki} -\CJ^{ij}   ~.}}}
\eqn\Noethercg{J_0^j= {1\over g^2} F_0^{\ j}\qquad , \qquad
J^{ij}= {1\over g^2 }F^{i j} -\CJ^{ij}~.}
Its conservation follows from the equation of motion of $A_j$ (the second in \fieldseomg).  Note that the antisymmetric part of $J^{ij}$ originates from the field strength and the symmetric part originates from the matter sector.

Unlike the current \vectorm\ in the matter theory, the current \Noethercg\ does not have explicit $x^i$ dependence.  As explains around \opssym\PVP, the reason for the explicit $x^i$ dependence in  \vectorm\ is that its charge does not commute with translations.  The commutator is a $U(1)$ transformation.  Here, that $U(1)$ transformation is a gauge transformation.  Therefore, the charge of \Noethercg\ commutes with translations on gauge invariant operators.  Hence, there is no explicit $x^i$ in \Noethercg.

The charge of the current \Noethercg\ acts only on the gauge field as
\eqn\vectoraction{A_i \to A_i -c_i~}
with constant $c_i$.  It leaves all other degrees of freedom invariant.  It is trivial to check that the Lagrangian \gaugeLag\ is invariant under \vectoraction.

So far space has been flat $\R^D$.  It is easy to generalize it.    For invariance of the Lagrangian $\CL_1$ \gaugeLag\ we need
\eqn\cigloc{\eqalign{
&\partial_0c_i=0\cr
&\partial_ic_j-\partial_jc_i=0\cr
&\grad^ic_i=0~.}}

This is reminiscent of a one-form global symmetry, where the action on the gauge field is also given by \vectoraction\ with $c_i$ a flat gauge field.
However, our $c_i$ is not a $U(1)$ gauge field.  This difference stems from the presence of the matter fields on which the symmetry \vectoraction\ does not act.  We might have attempted to identify two different $c_i$s that differ by $\partial_i\lambda$ by combining \vectoraction\ with a $U(1)$ gauge transformation by $\lambda$.  But this gauge transformation would transform the matter fields and will remain nontrivial.
A related fact is that $c_i$ must be single valued.  This is clear from its action on local operators.

Finally, as in the discussion around \Bmunucnv\Bmunugnv\ we can couple this system to gauge fields for this global symmetry.

\bigskip\centerline{\it A concrete realization}\bigskip

In order to find a concrete realization of this matter theory we deform the matter theory \Lzeron\ with its global symmetry \Cxiaction\ $\Phi \to e^{iC(x^i)}\Phi$, such that only the symmetry \opssym\ is present.  It acts as
\eqn\Cxiactions{\Phi \to e^{i\alpha +ic_ix^i}\Phi~.}
The resulting matter theory is essentially the one in \PretkoJBI.  Its leading order terms are
\eqn\Lzeroe{
\CL_0=i\bar\Phi \partial_0\Phi - s\partial_i(\bar\Phi\Phi)\partial^i(\bar \Phi\Phi)+ u \left(i\bar\Phi^2\partial_i\Phi\partial^i\Phi - i\Phi^2\partial_i\bar\Phi\partial^i\bar \Phi\right) -\lambda|\Phi|^4  +\cdots ~.}
The difference from \Lzeron\ is only in the term multiplying the real constant $u$.  This term breaks \Cxiaction\ to its subgroup \Cxiactions.

It is nice to write \Lzeroe\ in terms of $\Phi=\rho e^{i\theta}$
\eqn\effLagt{
\CL_0= -\rho^2\partial_0\theta   - s\partial_i(\rho^2)\partial^i(\rho^2)
- u \partial^i(\rho^4)\partial_i\theta  -\lambda\rho^4+\cdots~,}
where we dropped total derivatives.  This makes the global symmetries \Cxiactions, which act as $\theta \to \theta + \alpha + c_i x^i$, manifest.\foot{Motivated by this symmetry G.~Gabadadze suggested  a relation to Galileon \NicolisIN.}
However, it obscures the fact that the Lagrangian \Lzeroe\ is smooth at $\rho=0$.

The equation of motion of $\theta$ can be expressed as the condition \mattre\
\eqn\thetaeome{\eqalign{
&\partial_0\CJ_0-\partial_i\partial_j\CJ^{ij}=0\cr
&\CJ_0=-{\partial\CL_0\over \partial(\partial_0\theta)}= |\Phi|^2+\cdots \cr
&\CJ^{ij}= -\delta^{ij}u|\Phi|^4+\cdots}}
We see that our system has the two operators $\CJ_0$ and $\CJ^{ij}$ satisfying \mattre.  Therefore, this system is a special case of the matter theory discussed there and the entire discussion can be repeated.

Actually, the leading order Lagrangian \Lzeroe\effLagt\ leads to $\CJ^{ij}$ \thetaeome\ without the traceless symmetric tensor.  This fact is special to terms with up to two spatial derivatives.  These terms in the Lagrangian lead to terms without derivatives in $\CJ^{ij}$ and therefore cannot lead to nontrivial representations under rotations. Higher order terms can change that.  For example, the four-derivative term from \PretkoJBI\
\eqn\highero{v\Big|\Phi\partial_i\partial_j\Phi -\partial_i\Phi\partial_j\Phi\Big|^2=v\Big|\rho\partial_i\partial_j\rho-\partial_i\rho\partial_j\rho + i\rho^2\partial_i\partial_j\theta\Big|^2}
is manifestly invariant under \Cxiactions.  It leads to 
\eqn\hieCJ{\CJ^{ij}=\cdots +iv\Big(\bar\Phi^2(\Phi\partial_i\partial_j\Phi -\partial_i\Phi\partial_j\Phi)-\Phi^2(\bar\Phi\partial_i\partial_j\bar\Phi -\partial_i\bar\Phi\partial_j\bar\Phi)\Big)~.}
It includes both a trace and a traceless contribution.

The dipole operator $\CD_i=\bar\Phi \partial_i\Phi$ and the nonlocal operator $W( x, y)=\bar\Phi( x)\Phi( y)$ transform under \Cxiactions\ as (see also \transfoc)
\eqn\biltr{\eqalign{
&\CD_i=\bar\Phi \partial_i\Phi \to \CD_i +ic_i\bar\Phi\Phi \cr
&W( x, y)=\bar\Phi( x)\Phi( y) \to e^{i c_i ( y- x)^i}W( x, y)~.}}
This demonstrates again that this symmetry acts on local operators. 

The gauging of the symmetry associated with $\alpha$ in \Cxiactions\ is as in the discussion following \gaugeLag.  
This is achieved by changing all the derivatives to covariant derivatives $D_\mu \Phi =(\partial_\mu-iA_\mu)\Phi$ and $D_\mu \bar\Phi =(\partial_\mu+iA_\mu)\bar\Phi$ and using the gauge invariant field strength $F_{\mu\nu}=\partial_\mu A_\nu-\partial_\nu A_\mu$.

Then, \Lzeroe\ becomes (compare with \gaugeLag)
\eqn\Lonee{\eqalign{
\CL_{1}=&{1\over 4g^2} F_{\mu\nu}F^{\mu\nu} +i\bar\Phi D_0\Phi - s\partial_i(\bar\Phi\Phi)\partial^i(\bar \Phi\Phi)\cr
&+ u \left(i\bar\Phi^2\partial_i\Phi\partial^i\Phi - i\Phi^2\partial_i\bar\Phi\partial^i\bar \Phi + A_i \partial^i|\Phi|^4\right) -\lambda|\Phi|^4+\cdots}}
As above, we suppressed higher order terms and gauge invariant nonminimal couplings that are linear in $F_{\mu\nu}$.

The important term with coefficient $u$ can be written as
\eqn\termwithu{i\bar\Phi^2\partial_i\Phi\partial^i\Phi - i\Phi^2\partial_i\bar\Phi\partial^i\bar \Phi + A_i \partial^i|\Phi|^4 = i\bar\Phi^2D_i\Phi D^i\Phi - i\Phi^2D_i\bar\Phi D^i\bar \Phi~,}
thus making the $U(1)$ gauge symmetry manifest, but obscuring the fact that there is no ``seagull term'' proportional to $A_iA^i$ and that the term linear in $A_i$ multiplies a total derivative.

As in the general discussion around \vectoraction, the global symmetry associated with $c_i$ is simplified.  We can combine the action on $\Phi$ with a gauge transformation to make $\Phi$ invariant and instead transforming the gauge field
\eqn\oneform{\eqalign{
&\Phi\to \Phi\cr
&A_0\to A_0\cr
&A_i \to A_i - c_i~.}}

Adding the gauge fields to \biltr, the gauge invariant dipole operator $\CD_i=\bar\Phi D_i\Phi$ and the nonlocal operator $W( x, y)=\bar\Phi( x)e^{-i\int_{ x}^{ y}A}\Phi( y)$ transform under \oneform\ as
\eqn\biltrg{\eqalign{
&\CD_i=\bar\Phi D_i\Phi \to \CD_i +ic_i\bar\Phi\Phi \cr
&W( x, y)=\bar\Phi( x)e^{-i\int_{ x}^{ y}A}\Phi( y) \to e^{i c_i (y- x)^i}W( x, y)~.}}
We see again that this transformation acts on local operators.

Let us consider the theory on a spatial flat torus with radii $R_i$ and periodic boundary conditions.  When we studied the matter system with its global $U(1)$ symmetry, $\Phi$ had to be single valued and the global symmetry \Cxiactions\ was meaningful only when $c_i \in {1\over R_i}\Z$.  Now, when the global $U(1)$ is gauged, and we let the remaining global symmetry act only on the gauge field, as in \oneform, $c_i$ can be arbitrary real constants.  Note that the holonomies $e^{i\oint dx^j A_j}$ transform the same under $c_i $ and under $ c_i + {1\over R_i}$.  However, the transformations \biltrg\ show that these two values of $c_i$ cannot be identified.

\bigskip
\noindent {\bf Acknowledgments}

We are grateful to N.~Arkani-Hamed, M.~Cheng, G.~Gabadadze, Z.~Komargodski, H.T.~Lam, J.~Maldacena, M.~Pretko, S.-H.~Shao, S.~Sondhi, C.~Vafa, and A.~Vishwanath for helpful discussions.  We are particularly thankful to M.~Hermele for his patient explanations of fractons and for many useful comments on this approach.
This work was supported in part by DOE grant DE-SC0009988 and by the Simons Collaboration on Ultra-Quantum Matter, which is a grant from the Simons Foundation (651440, NS).
Opinions and conclusions expressed here are those of the authors and do not
necessarily reflect the views of funding agencies.

\appendix{A}{Canonical quantization and the spectrum of the matter theory}

Here we consider the matter theory $\CL_0$ \Lzeroe\ (or its special case with $u=0$ \Lzeron) and neglect higher order terms.  We stress that this is the theory before adding the $U(1)$ gauge field.

We start with the free theory, i.e.\ the terms with only two fields.  As we emphasized, it is important that these theories do not include the standard $|\partial_i\Phi|^2$ term.  This means that the only quadratic term in the Lagrangian is
\eqn\Lfree{\CL_{free}= i\bar\Phi \partial_0\Phi .}
Therefore, the energy is independent of the spatial momentum.  We interpret it to mean that the $\Phi$ ``particles'' have infinite mass.  Indeed, $\Phi$ quanta exist, but since \Lfree\ does not have spatial derivatives, they cannot move and hence they should not be called particles.

Let us analyze the theory in more detail.  In the quantum theory, the term \Lfree\ leads to
\eqn\commrel{[\Phi(x^i,x^0),\bar\Phi(y^i, x^0)]=\delta( x^i- y^i)~}
and therefore $\Phi$ includes annihilation operators and $\bar\Phi$ includes creation operators.

The Hamiltonian density derived from \Lzeroe\ is
\eqn\effH{
\CH_0= s\partial_i(\bar\Phi\Phi)\partial^i(\bar \Phi\Phi)- u \left(i\bar\Phi^2\partial_i\Phi\partial^i\Phi - i\Phi^2\partial_i\bar\Phi\partial^i\bar \Phi\right) +\lambda|\Phi|^4 ~}
and we take these terms to be normal ordered; i.e.\ all the $\Phi$s are to the right of all the $\bar \Phi$s.  Changing this ordering amounts to shifting $\CH_0$ by a linear combination of terms proportional to the unit operator (which shifts the overall energy of the system) and $|\Phi|^2$ (which can be removed by redefining $\Phi$).

We take the ground state $|0\rangle$ to be annihilated by the annihilation operators
\eqn\vacd{\Phi(x^i,x^0)|0\rangle =0 ~.}
Then, since every term in $\CH_0$ includes two annihilation operators, all the states
\eqn\xndef{
| x^i_1,x^i_2 , \cdots, x^i_n\rangle = \bar\Phi(x^i_1,x^0) \bar\Phi(x^i_2,x^0)\cdots \bar \Phi( x^i_n,x^0)|0\rangle \qquad {\rm with} \qquad x_l\not=x_m ~}
are annihilated by the Hamiltonian!

A more precise statement is that all the states \xndef\ with the same $n$ are degenerate.  States with different $n$ carry different $U(1)$ charges.  When we redefined $\Phi$ and removed the quadratic term $|\Phi|^2$ from the Lagrangian, we shifted $\CH_0 $ by a term proportional to the $U(1)$ charge density $J_0=|\Phi|^2+\cdots$ \thetaeome.  Only after this shift are the states \xndef\ with different $n$  degenerate.

It is important to emphasize that the coupling of the system to a dynamical $U(1)$ gauge field, as we do in sections 3 and 4, removes this huge degeneracy.  These states do not satisfy Gauss law.

\listrefs

\end